\documentclass{birkjour}
\usepackage[extra]{tipa}
\usepackage{url}
\usepackage[noadjust]{cite}
\usepackage{enumerate}
\newcommand{\be}{\begin{equation}}
\newcommand{\ee}{\end{equation}}
\newcommand{\BC}{\mathbb{C}}
\newcommand{\BR}{\mathbb{R}}

\newcommand{\cC}{\mathcal{C}}
\newcommand{\cD}{\mathcal{D}}

\newcommand{\ed}{\end{document}}

 \newtheorem{thm}{Theorem}[section]
 \newtheorem{rem}[thm]{Remark}
\begin{document}
\date{}
\title{START in a five--dimensional conformal domain}
\author{Arkadiusz Jadczyk}
\address{Center CAIROS, Institut de Math\'{e}matiques de
    Toulouse\\Universit\'{e} Paul Sabatier, 31062 TOULOUSE CEDEX 9, France}
\email{arkadiusz.jadczyk@cict.fr}
\begin{abstract}
In this paper we give a brief review of the pseudo--Riemannian geometry of the five--dimensional homogeneous space for the conformal group O(4,2). Its topology is described and its relation to the conformally compactified Minkowski space is discussed. Its metric is calculated using a generalized half--space representation. Compactification via Lie-sphere geometry is outlined. Possible applications to Jaime Keller's START theory may follow by using its predecessor - the 5-optics of Yu. B. Rumer. The point of view of Rumer is given extensively in the last section of the paper
\end{abstract}
\thanks{This paper is dedicated to the memory of Jaime Keller }
\date{}
\maketitle
\section{Introduction}
Jaime Keller \cite{keller1}  (cf. also the detailed analysis by da Rocha \cite{darocha1}) expressed conformal transformations via the adjoint representation of $Spin(4,2)$ acting on paravectors of the Clifford algebra $Cl_{4,1}$ and twistors as elements of a left minimal ideal of the Dirac-Clifford algebra $\BC\otimes Cl_{1,3}.$ This observation possibly contributed to his interest in a five--dimensional formulation of physics, where, following Yu. B. Rumer (cf. \cite{rumer1,rumer2} and references therein), he gave the interpretation of the fifth coordinate as related to {\it action\,}. He has elaborated this idea in a sequence of papers using the name START (Space-Time-Action Relativity Theory) - cf. e.g. \cite{keller2,keller3,keller4}.\\
In the present paper we study some geometrical properties of two five--dimensional homogeneous domains for the conformal group $SO(4,2)$\footnote{In the mathematical literature one can find several slightly different definitions of the ``conformal group'' of a pseudo--Euclidean space $\BR^{r,s}$. Some authors define it as $SO(r+1,s+1),$ some other as $PO(r+1,s+1).$ We choose simply $O(r+1,s+1)$ because in our construction of the double cover of the compactified Minkowski space $\doubletilde{M}$ inversions play a nontrivial role.}, with the hope that these properties may prove to be relevant for further extension of the ideas developed in Keller's START theory. We will also discuss some of the ideas of Keller's predecessor, Yu. B. Rumer, whose works have been published only in Russian, and therefore are largely unknown for the rest of the world.
\section{Conformally compactified Minkowski space as a boundary of five--dimensional domains}
Conformally compactified Minkowski space is the Shilov's boundary of an eight--dimensional complex domain $$SO(4,2)/S(O(4)\times O(2))\approx SU(2,2)/S(U(2)\times U(2))$$ - cf. \cite{jadczyk1} and references therein. But it can also be considered as a boundary of a five--dimensional homogeneous space for the conformal group. We will now study this latter case in some detail.

\smallskip
We denote by $\BR^{r,s}$ the space vector space $\BR^n,\, n=r+s$ endowed with the quadratic form \be q(x)=(x^1)^2+...+(x^r)^2-(x^{r+1})^2-...-(x^{r+s})^2.\ee When $s=0$ (or when $r=0$) the natural compactification of $\BR^{r,s}=\BR^n$ is the Alexandroff's one-point compactification, that is the $n$-sphere $S^n.$ In other cases the natural compactification is the so-called {\em conformal compactification\,}. For the Minkowski space, $r=3,s=1$ (or $r=1,s=3$) the resulting compactified space consists of the Minkowski space with added {\em conformal infinity\,} - a three--dimensional variety of a generalized Dupin cyclide, sometimes misleadingly called "the light cone at infinity" - cf. \cite{jadczyk1,jadczyk2} and references therein.

\subsection{Compactified Minkowski space}
As we will see, compactified Minkowski space can be viewed as a four--dimensional boundary between two five--dimensional domains. Boundaries and regions close to boundaries are interesting. We can easily imagine that the physical Reality is five--dimensional and that, for reasons yet to be understood, our perception is restricted to a thin four--dimensional boundary. While physics\footnote{In fact, since we are dealing here with human perception, an interdisciplinary approach is needed, including the theory of information, biology, and even some philosophy.} of such an approach may be still in a development (as in Kaluza-Klein type theories or in Jaime Keller's unfinished START program), mathematics is not that difficult and we will describe it briefly in the following.
\smallskip
Consider $\BR^{4,2}$ endowed with coordinates $X^1,...,X^6,$ scalar product $(X,Y)=X^1Y^1+...+X^4Y^4-X^5Y^5-X^6Y^6,$  and the quadratic form $Q(X)=(X,X).$ There we have a null cone consisting of those $X$ for which $(X,X)=0.$ This null cone is singular - it has an apex at $X=0$ and it is useful to remove this apex. Let us denote the remaining set by $\cC:$
\be \cC=\{X\in\BR^{4,2}:\, (X,X)=0,\, X\neq 0\}.\ee
The standard way is to consider the set of all generator line of the cone $Q(X)=(X,X)=0$ or, what is the same, to divide $\cC$ by the equivalence relation $X\sim Y$ if and only if $X=cY$ for some $c\in \BR$ - then necessarily $c\neq 0.$ The resulting set (a {\it projective quadric\,}) denoted $\tilde{M},$ $\tilde{M}=\cC/\sim,$ happens to be a four--dimensional compact manifold, diffeomorphic to $(S^3\times S^1)/Z_2$ - the compactified Minkowski space. But there is another option: instead of taking the quotient by real numbers, we can divide $\cC$ by a stronger equivalence relation, namely $X\approx Y$ if and only if $X=cY, c>0.$ The resulting manifold $\doubletilde{M}=\cC/\approx$ is a double covering of $\tilde M.$ \footnote{While physics of such a construction is speculative, mathematically this second construction is not less natural than the standard one.}

\smallskip
We can now embed Minkowski space $M=\BR^{3,1}$, with coordinates $(x^\mu)=(x^1,...,x^4),$ and with the quadratic form \be q(x)=(x^1)^2+...+(x^3)^2-(x^4)^2,\ee
 using a variation of the standard formula (cf. \cite[p. 80, (B)]{angles} and \cite[Eq. (9)]{jadczyk1}):
\be \tau(x)=(x,\frac{1}{2}(1-q(x)),-\frac{1}{2}(1+q(x)).\ee
It can be easily verified that $Q(\tau(x))=0,$ thus $\tau(x)\in \cC$, and that $X\in\cC$ is in $\tau(M)$ if and only if $X^5-X^6=1.$ The remaining part of $\cC,$ namely the part characterized by the condition $X^5=X^6,$ when divided by the equivalence relation $\sim,$ projects onto {\em conformal infinity}. \\
With the stronger equivalence relation $\approx$ we have two non intersecting embeddings of $M$ into $\doubletilde{M}$ described by:
\be \tau_+(x)=(x,\frac{1}{2}(1-q(x)),-\frac{1}{2}(1+q(x))).\label{eq:taup}\ee
\be \tau_-(x)=(x,-\frac{1}{2}(1-q(x)),\frac{1}{2}(1+q(x))).\label{eq:taum}\ee\\
Compactified Minkowski space, being a projection of the null cone $\cC$ does not inherit from the quadratic form $Q$ of $\BR^{4,2}$ any natural pseudo-Riemannian structure. It inherits only a conformal structure (of signature $(3,1)$) - cf. \cite{jadczyk1}.
\subsection{Compactified Minkowski space as a boundary}
The null cone $\cC$ of $\BR^{4,2}$ separates two domains $D_\pm$ characterized by the condition
\be  D_+=\{X\in\BR^{4,2}:\,Q(X)>0\},\, D_-=\{X\in\BR^{4,2}:\,Q(X)<0\}.\ee
Let us consider their projections $\cD_+$ (resp. $\cD_-$) obtained by taking the quotient by the equivalence relation $\approx.$ For every point of $\cD_+$ (resp. $\cD_-$) there is a unique point $X$ in $D_+$ (resp. $D_-$ for which $Q(X)=1$ (resp. $Q(X)=-1).$ Therefore $\cD_\pm$ can be identified with the hyperboloid $\Sigma_\pm$ defined by
\be \Sigma_{\pm}=\{X\in\BR^{4,2}:\,Q(X)=\pm 1\}.\ee
Now the quadratic form $Q$ defines a pseudo-Riemannian metric on $\Sigma_\pm.$ In fact we have the following theorem \cite[p. 66]{wolf}\footnote{Thanks are due to Pierre Angl\`{e}s for bringing this reference to author's attention.}
\begin{thm}$\Sigma_+$ (resp. $\Sigma_-$) is a complete pseudo--Riemannian manifold of constant curvature, of signature $(3,2)$ (resp. $(4,1)$). The geodesics of $\Sigma_\pm$ are intersections $P\cap\Sigma_\pm$ of $\Sigma_\pm$ with planes $P$ through $0$ in $\BR^{4,2}.$ The group of all isometries of $\Sigma_\pm$ is $O(4,2).$
\end{thm}
So, in the projective space $\mathbf{P}(\BR^{4,2})$ we have two five--dimensional pseudo-Riemannian manifolds of signatures $(3,2)$ and $(4,1)$ respectively, separated by a compact four--dimensional manifold endowed with a conformal structure only, of signature $(3,1)$ - the compactified Minkowski space.\\
Let us look now at the topology of the two five--dimensional domains $\Sigma_\pm.$ For the domain $\Sigma_+$ we have the defining equation
\be (X^1)^2+...+(X^3)^2-(X^4)^2+(X^5)^2-(X^6)^2=1. \label{eq:smp}\ee We can write it as $$(X^1)^2+...+(X^3)^2+(X^5)^2=(X^4)^2+(X^6)^2+1.$$ It is then clear that $X^4$ and $X^6$ can be arbitrary real numbers, and that introducing $Y^i=X^i/((X^4)^2+(X^6)^2+1),\, (i=1,2,3,5)$ we have $(Y^1)^2+...(Y^3)^2+(Y^5)^2=1.$ Therefore $\Sigma_+$ has the topology of $S^3\times\BR^2.$ Using a similar reasoning we easily deduce that $\Sigma_-,$ defined by the condition
\be (X^1)^2+...+(X^3)^2-(X^4)^2+(X^5)^2-(X^6)^2=-1\label{eq:smm}\ee has the topology of $S^1\times\BR^4.$

\section{An explicit description of the domain $\Sigma_-$}
In this section we will introduce a particular set of local coordinates in $\Sigma_-$ and calculate an explicit expression for the induced metric. To this end will adapt the method discussed by Cannon et al. \cite[Chapter 7]{cannon} who, discussing the ``Five Models of Hyperbolic Space'', similarly to Wolf \cite[p. 70]{wolf}, considers only the hyperbolic case of $\BR^{n-1,1}.$ Our case of $\BR^{4,2}$ is somewhat more singular, thus some care needs to be taken, but otherwise the formal reasoning is similar.
\subsection{Local coordinates and the metric}
In $\Sigma_-$ we choose an open set defined by the condition $X^5-X^6>0.$ On this set we introduce five coordinates  $(x^\mu,\lambda)\in\BR^5$ defined by
\be x^\mu=\frac{X^\mu}{X^5-X^6}\,,\lambda=\frac{1}{X^5-X^6}>0.\label{eq:coor1}\ee
On the other hand, given a point in $\BR^5$ with coordinates $(x^\mu,\lambda>0)$ we can embed it  in $\Sigma_-$ as follows:
\be (X^\mu)=\frac{x^\mu}{\lambda},\, X^5=\frac{1-q(x)-\lambda^2}{2\lambda},\,X^6=-\frac{1+q(x)+\lambda^2}{2\lambda}.\label{eq:coor2}\ee
The Reader is encouraged to verify by a straightforward calculation that, with the above definition, $Q(X(x^\mu,\lambda))=-1,$ and that  applying the formula (\ref{eq:coor1}) to $X(x^\mu,\lambda)$ we indeed recover $(x^\mu,\lambda).$
\smallskip
We can now calculate the metric. In general, when we are dealing with an embedded manifold parameterized by coordinates $x^\alpha,$ the metric $g_{\alpha,\beta}$ induced from the metric $G_{AB}$ in which our manifold is embedded of is given by the expression
\be g_{\alpha\beta}=\frac{\partial X^A}{\partial x^\alpha}\frac{\partial X^B}{\partial x^\beta}\,G_{AB}.\ee

In our case $(G_{AB})=\mbox{diag }(1,1,1,-1,1,-1)$ and it is easy to calculate $g_{\alpha\beta}$ using the formula (\ref{eq:coor2}). The result of a straightforward calculation is:
\be (g_{\alpha\beta})=\frac{1}{\lambda^2}\mbox{diag }(1,1,1,-1,1).\label{eq:metric}\ee
Exactly the same method applies to the region $X^5-X^6<0.$ We get a five--dimensional pseudo--Riemannian, conformally flat, manifold of constant curvature and signature $(4,1).$ We have covered by coordinates two regions corresponding to different signs of the fifth coordinate. Physicists, when discussing representations of the conformal group with applications to elementary particle physics,often restrict their attention to these regions - cf. for instance \cite{ingraham1,ingraham2,ingraham3}. Yet evidently the group $O(4,2)$ acts on this part with singularities. Like in the case of Minkowski space in order to avoid singularities one has to add ``conformal infinity''. In our case this is the region where $X^5=X^6.$ This conformal infinity of the five--dimensional domain has a simpler structure than the one for Minkowski space. In fact, setting $X^5=X^6$ in (\ref{eq:smm}) we get $$ (X^1)^2+...+(X^3)^2-(X^4)^2=-1$$ with no scaling freedom. \\Therefore the conformal infinity of our five--dimensional domain $\Sigma_-$ is the Cartesian product of $\BR$ ($X^5-X^6\in\BR$) and the standard two-sheeted hyperboloid of Minkowski's space.
\subsection{Christoffel symbols and geodesics}
Given the metric (\ref{eq:metric})  it is easy (in our coordinate patch) to calculate the Christoffel symbols ${\Gamma^\mu}_{\nu\sigma}$ and geodesic equations - cf. e.g. \cite[Mathematica Programs: Christoffel Symbols and Geodesic Equations]{hartle}. The metric is conformally flat and the only non--vanishing Christoffel symbols are:
\begin{align}
{\Gamma^\mu}_{5\sigma}&=-\frac{1}{\lambda}\,\delta^\mu_\sigma\\
{\Gamma^5}_{\nu\sigma}&=\frac{1}{\lambda}\,\eta_{\nu\sigma}\\
{\Gamma^5}_{55}&=-\frac{1}{\lambda}.
\end{align}
The corresponding geodesic equations, when parametrized by an affine parameter $s$, are:
\begin{align}
\frac{d^2x^\mu}{ds^2}&=\frac{2}{\lambda}\,\frac{dx^\mu}{ds}\frac{d\lambda}{ds}\\
\frac{d^2\lambda}{ds^2}&=-\frac{1}{\lambda}\,\left((\frac{dx^1}{ds})^2+(\frac{dx^2}{ds})^2+(\frac{dx^3}{ds})^2-(\frac{dx^4}{ds})^2-(\frac{d\lambda}{ds})^2\right),
\end{align}
where $\mu,\nu,\sigma=1,..,4,$ and $\eta_{\nu\sigma}$ is the flat Minkowski metric $\mbox{diag}(1,1,1,-1).$
It is interesting to notice that, for $\lambda=\mbox{const.},$ Minkowski's space null lines $x^\mu (s)=su^\mu,$ where $u^\mu,\,\mu=(1,...,4) $ is a fixed null vector, are geodesics of the five--dimensional space.\\
\noindent When $\lambda$ is non--constant, it is convenient to choose $\lambda$ as a (non--affine) parameter. The geodesic equations will read in such a case (adapted from \cite[Appendix B, (B7)]{muller}):
\begin{align} 0=&\frac{d^2x^\mu}{d\lambda^2}-\frac{dx^\mu}{d\lambda}\left({\Gamma^5}_{55}+2{\Gamma^5}_{5\nu}\,\frac{dx^\nu}{d\lambda}+{\Gamma^5}_{\nu\sigma}\frac{dx^\nu}
{d\lambda}\frac{dx^\sigma}{d\lambda}\right)\nonumber\\
&+{\Gamma^\mu}_{55}+2{\Gamma^\mu}_{5\nu}\frac{dx^\nu}{d\lambda}+{\Gamma^\mu}_{\nu\sigma}\frac{dx^\nu}{d\lambda}\frac{dx^\sigma}{d\lambda},
\end{align}
which, in our case, reduce to:
\be
x''^\mu(\lambda)=x'^\mu(\lambda)\frac{1+x'^2(\lambda)}{\lambda},
\ee
where we denoted by a prime the derivative with respect to $\lambda,$ and used the notation $x'^2(\lambda)=\eta_{\nu\sigma}\frac{dx^\nu}{d\lambda}\frac{dx^\sigma}{d\lambda}.$
The direction of the vector $x'^\mu$ is kept constant along the geodesics. Thus we need to consider three cases: $x'^2=0, x'^2<0,x'^2>0.$ If $x'^2=0,$ we can use a Lorentz rotation
(in the variables $x^\mu$) to set the direction of $x'^\mu$ along the vector $(1,0,0,1).$   The differential equations reduce in this case to the following ones:
\be {x^1}''(\lambda)={x^1}'(\lambda)/\lambda,\,   {x^4}''(\lambda)={x^4}'(\lambda)/\lambda,\ee
which, taking into account the constraint ${x'}^2=0,$  solve to
\be x^1(\lambda)=a\lambda^2+x^1_0,\, x^4(\lambda)=a\lambda^2+x^4_0,\ee
with $x^2$ and $x^3$ constant.\\
When $x'^2<0,$ we can use a Lorentz rotation to rotate the geodesic into $(x^4,\lambda)$ plane. The relevant differential equation:
\be {x^4}''(\lambda)=(1-{x^4}'(\lambda)^2)/\lambda\ee
solves to $(x^4(\lambda)-x^4_0)^2-\lambda^2=a^2$ - a hyperbola.\\
When ${x'}^2>0,$ we can Lorentz rotate the geodesic into $(x^1,\lambda)$ plane, and the differential equation
\be {x^1}''(\lambda)=(1+{x^1}'(\lambda)^2)/\lambda \ee solves to $(x^1(\lambda)-x^1_0)^2+\lambda^2=a^2$ - a semi--circle.

\begin{figure}[!ht]
\begin{center} \includegraphics[width=9cm, keepaspectratio=true,clip]{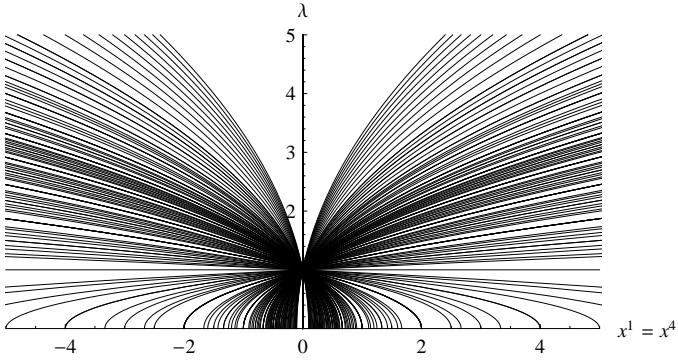}
\end{center}
\caption{A family of geodesics in $(x^1=x^4,\lambda)$ plane through the point $(0,1).$}
\label{fig:fig1}
\end{figure}

\begin{figure}[!ht]
\begin{center} \includegraphics[width=9cm,
keepaspectratio=true,clip]{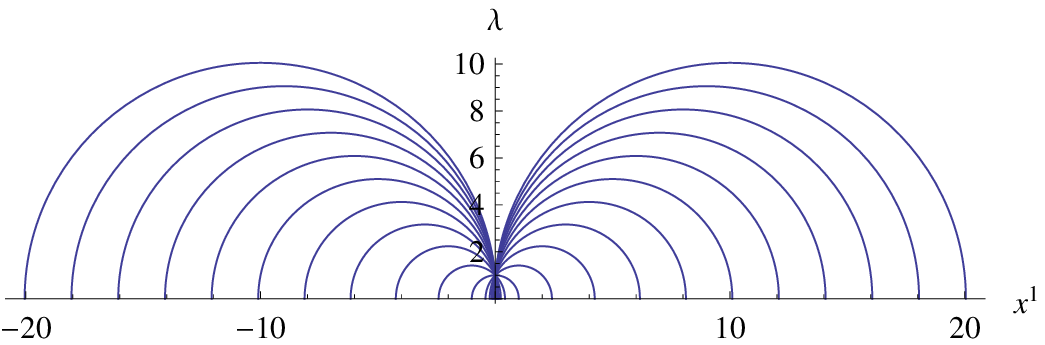}
\end{center}
 \caption{A family of geodesics in $(x^1,\lambda)$ plane through the point $(0,1)$.}
\label{fig:fig2}
\end{figure}

\begin{figure}[!ht]
\begin{center} \includegraphics[width=9cm,
keepaspectratio=true,clip]{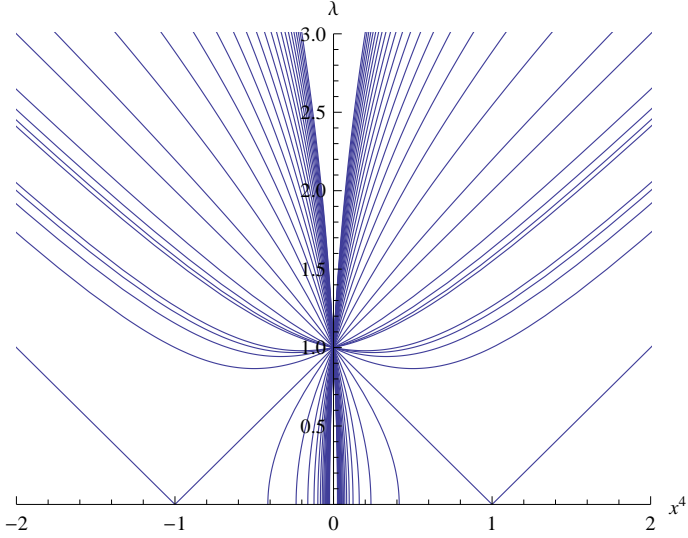}
\end{center}
 \caption{A family of geodesics in $(x^4,\lambda)$ plane through the point $(0,1)$.}
\label{fig:fig3}
\end{figure}

\subsection{$\Sigma_-$ as the space of hyperboloids}
The five--dimensional homogeneous space $\Sigma_-$ can be interpreted as the space of (unoriented) hyperboloids in Minkowski space, along the lines of a generalized M\"{o}bius geometry (cf. e.g. \cite[Ch. 1.2]{cecil1}). Let $Y$ be in $\BR^{4,2}$ with $Y^5-Y^6>0$ and $Q(Y)<0.$ Consider the set of all $x\in M$ for which $(\tau_+(x),Y)=0.$ Normalizing $Y$ so that $Y^5-Y^6=1$ we can write it in the form $Y=\left(y^\mu,\frac{1-q(y)-\lambda^2}{2},-\frac{1+q(y)+\lambda^2}{2}\right).$ A simple calculations shows that the condition $(\tau_+(x),Y)=0$ translates then to $q(x-y)=-\lambda^2.$ For $y=0$ this is a double-sheeted hyperboloid with apex at $x^1=x^2=x^3=0,x^4=\pm \lambda.$ Each geodesic line in $\Sigma_-$ can thus be interpreted as a particular one--parameter family of hyperboloids in Minkowski space.

\subsection{The case of $\Sigma_+$}
The same method applies as above applies in this case, except that there is a change of signs in front of $\lambda^2$ in (\ref{eq:coor2}). The resulting metric is then
\be (g_{\alpha\beta})=\frac{1}{\lambda^2}\,\mbox{diag }(1,1,1,-1,-1),\ee
with signature $(3,2).$ As in the case of $\Sigma_-$ the conformal infinity is the Cartesian product of $\BR$ and, this time, the one-sheeted hyperboloid $$ (X^1)^2+...+(X^3)^2-(X^4)^2=1$$
Minkowski space can be embedded in our five--dimensional manifold simply by putting $\lambda=1.$  It follows that the direction of the vector $x'$ is constant along geodesics.
\begin{rem}
Following Wolf \cite{wolf} we have considered in details only the case of the equivalence relation $\approx .$ In projective geometry one is using the weaker relation $\sim.$ The standard projection can be discussed along the same lines as above. In that case the regions $X^5-X^6>0$ and $X^5-X^6<0$ are identified, so we can restrict our attention to $\lambda>0.$\footnote{That is why a similar coordinatization is often referred to as ``half--space model'' in the literature on hyperbolic geometry - see e.g. \cite{elstrodt}. } On the other hand, when discussing the topology - we have to additionally take the quotient by $Z_2.$
\end{rem}
\subsection{Compactification of the five--dimensional space}
The manifolds $\Sigma_\pm,$ having the topology of $S^2\times\BR^3$ and $S^1\times\BR^4,$ are non--compact. They relate to M\"{o}bius geometry of hyperboloids in Minkowski space. Yet there is an extension of M\"{o}bius geometry - the geometry of Lie Spheres, initiated by Sophus Lie \cite{lie1} and developed by  Wilhelm Blaschke \cite{blaschke3} and Thomas E. Cecil \cite{cecil1}. Using Lie spheres approach we would end up with a projective null cone in a seven--dimensional space that would lead to a compactified version of $\Sigma_\pm.$ These compact versions would be equipped with $O(4,3)$ (resp. $O(5,2)$) invariant conformal structures. When restricting the symmetry group to $O(4,2)$ we would then obtain the compactified versions of our pseudo--Riemannian five--dimensional manifolds $\Sigma_\pm.$ Yet, as of today, this line of research seems to be unfinished.\footnote{ For a rough justification of such an approach cf. e.g. \cite{wyler68,wyler-IAS}.}
\section{Predecessor: 5-optics of Yu. B. Rumer}
In his 2002 paper \cite{keller3} Jaime Keller gives references  to five papers of Albert Einstein, seven papers of his own, and two references to Yu. B. Rumer \cite{rumer1,rumer2}. In a long series of papers (years 1949--1959)  Yu. B. Rumer developed a five--dimensional formulation of physics, extending the early ideas of Kaluza, Klein, Einstein, Bergmann and Bargmann. Jaime Keller makes in \cite{keller3} the following remarks:
\begin{quotation}Besides the many papers which have been written about the Kaluza-Klein proposition and
their extension to the idea of hyper-space with one additional dimension (at least) for each
additional interaction included, the direct inclusion of action as a fifth dimension was proposed
as early as the 1949-1956 by the Russian physicist Y.B. Rumer [13, 14] under the name of ``Action as a spatial coordinate. I--X''. In the work of Rumer the main foreseen application is to
the case of optics in what he called 5-optics. We should remember that in this case the action
$dA = 0$ and then the fifth coordinate turns out to be identically null.\end{quotation}
While the meaning of last part (``the fifth coordinate turns out to be identically null") is unclear to the present author, it should be stressed that Rumer applied his methods to more than the case of optics - although geometrical optics was his starting point. It may be instructive to recall Rumer's own comments on his theory - these comments, from the afterword to his 1956 monograph \cite{rumer1} give a historical overview of the involved ideas.
\smallskip
Rumer mentions that the formal apparatus of his ``5-optics" was essentially built in the works of Theodor Kaluza, Oskar  Klein, Vladimir Fock, Albert Einstein and Peter Bergmann. Then he gives an historical overview of five--dimensional theories with these remarks:
\begin{enumerate}[I]
  \item Kaluza (1921) \cite{kaluza}
  \begin{enumerate}[1]
  \item The extra fifth dimension of the four--dimensional physical space of the theory of gravity is introduced. Physical meaning of extra dimension remains open.
  \item It is realized that the metric potentials of the 5-space should not depend on the extra fifth coordinate. A physical meaning of this condition remains open.
  \item In order to have a possibility of a one-to-one correspondence between 10+4=14 potentials of the theory of gravity and electrodynamics with 15 metric potentials of the 5--space, an additional condition is being introduced, namely that $G_{55}=1.$ Physical meaning of this requirement remains open.
  \end{enumerate}
  \item O. Klein \cite{klein} and V. A. Fock (1926) \cite{fock}
  \begin{enumerate}[1]
  \item The relation between 14 potentials of the theory of gravity and electrodynamics, and 15 metric potentials of the 5-space is made more precise. A trajectory of a charged particle is described as a null geodesic line (geometrical ray) in 5-space. In fact one gets an equivalence of the problem of relativistic classical mechanics of a motion of a material point with the problem of geometrical optics of the ray propagation in 5-space.
  \item It is found that there is a possibility of a formulation of the quantum mechanical problem of the motion of a charged particle as a problem of the wave optics of scalar waves propagation in 5-space, namely if one imposes periodicity condition:
      $$W(x^1,x^2,x^3,x^4,x^5)=U(x^1,x^2,x^3,x^4)\exp \left(i(\frac{mc}{\hbar})x^5\right),$$
      while keeping the cylindricity condition for the metric potentials.
  \item The problems of a physical meaning of the fifth coordinate, the cylindricity condition for the metric potentials, and periodicity condition for the wave function remain open. The question about a physical meaning of the condition $G_{55}=1$ remains open
  \end{enumerate}
  \item A. Einstein and P. Bergmann (1938) \cite{einstein-bergmann}
The condition of cylindricity is replaced by a weaker condition of periodicity of  metric potentials in the fifth coordinate. The period is assumed to be of a microscopic dimension, which, in a first approximation, can be put equal to zero. In such a case the periodicity condition degenerates into that of cylindricity.

As there is no equivalence condition for the electromagnetic field, in all these papers the metric tensor of the 5--space depends  on the ratio $\frac{e}{m}$ for the particle of which motion is being considered, while the metric tensor for the 4--space is a universal one.

From this fact one has to deduce that the 5--space of five--dimensio\-nal generalizations of the theory of gravity  cannot be (extended by one extra dimension) a universal physical space of general relativity, but it should have some different physical meaning.
    \item 5-optics
    \begin{enumerate}[1]
\item The 5--space of 5--optics is  a (extended by one extra dimension) configuration space for the test particle under consideration. Metric and topological structure of this space reflects the character of action of all external matter  on the particle.
    \item  The fifth coordinate of the configuration space has a clear physical character of action. The 5-space is closed in the fifth coordinate.
    \item Instead of the condition of cylindricity for metric potentials and cyclicity for the wave function all physical quantities are subjected to just one condition of periodicity in the 5-th coordinate.
        \item One finds that the period of the fifth coordinate has a universal value of Planck's constant, what has a clear physical meaning.
    \item Quantization of action of a material point is an effect of a periodical dependence of physical quantities on the action coordinate.
    \item The possibility of assuming $G_{55}=1$ in the previous theories is conditioned by the fact that the 5-eiconal equation
    $$G^{\mu\nu}\,\frac{\partial \Sigma}{\partial x^\mu}\frac{\partial \Sigma}{\partial x^\nu}=0,$$
    which formulates the classical mechanical problem of motion of a charged material point, is {\em homogeneous\,} in metric potentials $G^{\mu\nu}.$ Therefore, in this problem, only fourteen relations between metric potentials have a physical meaning, and the condition $G_{55}=1$ does not lead to a contradiction.
    \item A different situation arises in the problem of defining metric potentials from given external sources of the field, which is formulated through Einstein's equations for 5--space:
    $$P_{\lambda\mu}-\frac12\,G_{\lambda\mu}P=\kappa\,Q_{\lambda\mu}.$$
    which are inhomogeneous in metric potentials.

    When solving this problem the ratio $\frac{e}{m},$ that enters the expression for the metric tensor for 5-space, should be replaced by a universal quantity $c^2\sqrt{\frac{\kappa}{2\pi}}.$ The value of the potential $G_{55}$ should then be derived from field equations. One should not put $G_{55}=1$ from the beginning, as this leads, for instance, to wrong results in the problem of a charged point mass.
    \item Taking into account the periodical dependence of the electromagnetic field from the fifth coordinate leads automatically not only to long range Coulomb--type forces, but also to short--range forces of Yukawa's type.
        \item In all resulting classical theory we are obliged to take $\hbar\rightarrow 0,$ i.e. we should neglect the periodical dependence of physical quantities on the action coordinate. In all resulting quantum theory we are obliged to take into account periodical dependence of physical quantities on the action coordinate. Therefore, from the point of view 5--optics, it is inappropriate to neglect, as it is in classical mechanics, the periodical dependence of the components of the external field on the action coordinate.

Taking into account this dependence should lead to a prediction and discovery of a number of 5-optics effects, which could be then used for an experimental verification of the theory.
  \end{enumerate}
\end{enumerate}

In Rumer's autobiographical notes \cite{plastinki} he mentions that his theory of 1956 was predicting spin $3/2$ for the electron - three times too much. This fact has discouraged Freeman Dyson who, for some time, was following all developments of Rumer's theory. It is only in 1959, with the help of V. Pokrovsky, Rumer was able to find a solution to this puzzle \cite{rumer2}.
\section*{Acknowledgment}Thanks are due to Pierre Angl\`{e}s for his interest, comments and encouragement, and to Rafal Ablamowicz for his  supervision and help in the editorial issues. Support of Quantum Future Group (QFG) is warmly acknowledged.

\end{document}